\documentclass[12pt,epsf,epsfig,psfig]{article}
\usepackage{graphicx}
\usepackage{epstopdf}
\usepackage{epsfig}
\usepackage{cite}
\def\e{\epsilon}

\def\B{\boldmath}

\def\be{\begin{equation}}
\def\ee{\end{equation}}

\def\lsim{\raise0.3ex\hbox{$<$\kern-0.75em\raise-1.1ex\hbox{$\sim$}}}
\def\gsim{\raise0.3ex\hbox{$>$\kern-0.75em\raise-1.1ex\hbox{$\sim$}}}

\def\NP{{ Nucl.\ Phys.\ }}
\def\PL{{ Phys.\ Lett.\ }}
\def\PR{{ Phys.\ Rev.\ }}

\def\PRL{{ Phys.\ Rev.\ Lett.\ }}

\begin{document}



December 2015 \hfill BI-TP 2015/20

~~\vskip0.3cm

\centerline{\Large \bf Strangeness Production}

\medskip

\centerline{\Large \bf in \B$AA$ and \B$pp$ Collisions}

\vskip0.3cm 

\centerline{\bf Paolo Castorina$^{\rm a}$
and Helmut Satz$^{\rm b}$} 

\bigskip

\centerline{a: Dipartimento di Fisica ed Astronomia, 
Universita' di Catania,} 

\centerline{
and INFN, Sezione de Catania, Catania, Italy}

\centerline{b: Fakult\"at f\"ur Physik, Universit\"at Bielefeld, Germany}



\vskip0.5cm

\centerline{\large \bf Abstract}

\bigskip

Boost-invariant hadron production in high energy collisions occurs in 
causally disconnected regions of finite space-time size. As a result, 
globally conserved quantum numbers (charge, strangeness, baryon number) 
are conserved locally in spatially restricted correlation clusters. 
Their size is determined by two time scales: the equilibration time 
specifying the formation of a quark-gluon plasma, and the hadronization 
time, specifying the onset of confinement.
The expected values for these scales provide the theoretical basis for 
the suppression observed for strangeness production in elementary interactions 
($pp$, $e^+e^-$) below LHC energies. In contrast, the space-time superposition 
of individual collisions in high energy heavy ion interactions leads to higher 
energy densities, resulting in much later hadronization and hence much larger
hadronization volumes. This largely removes 
the causality constraints and results in an ideal hadronic resonance gas in 
full chemical equilibrium. In the present paper, we determine the collision 
energies needed for that; we also estimate when $pp$ collisions reach 
comparable hadronization volumes and thus determine when strangeness 
suppression should disappear there as well. 

\vskip1.5cm

The main aim in the study of high energy nucleus-nucleus collisions is the
production of the quark-gluon plasma predicted as a new state of matter by
statistical QCD. It is expected that nuclear collisions, in contrast to
proton-proton interactions, will provide a much larger interaction volume
and thus allow an investigation of bulk QCD features. The first step in
such a study is therefore the determination of observables which show a 
different behavior in $pp$ and in $AA$ collisions. The three basic deviations 
in $AA$ collisions so far established are 

-- enhanced strangeness production \cite{MR},

-- supppressed quarkonium production \cite{MS}, and

-- jet quenching \cite{Bj,GW} 

Our topic in the present paper is the issue of strangeness production. In 
$pp$ collisions, one finds a suppression of strange hadron production 
relative to an equilibrium distribution of species abundances at the 
hadronization temperature. In nuclear collisions at sufficiently high 
energy (RHIC and above), this suppression disappears; now also strange 
hadrons are produced in accord with the ratios predicted by a grand 
canonical resonance gas at $T_H$.

\medskip

The strangeness suppression in $pp$ collisions as well as in low energy
nuclear collisions has been accounted for in terms of local strangeness
conservation \cite{Red1}. To conserve strangeness, a produced $s$ quark
has to be compensated by a corresponding $\bar s$ sufficiently nearby. 
If in a given rapidity range only a single strange particle pair is
produced, the use of an equivalent overall composition volume \cite{beca}
for a resonance gas is not valid. Strangeness conservation then requires a
smaller conservation volume, and this leads to an effective reduction of
the strangeness production rates. In high energy nuclear collisions, the
superposition of many nucleon-nucleon interaction volumes leads to 
abundant strangeness production and thus removes the need for a smaller
conservation volume.

\medskip

In a recent paper \cite{CS-c}, we had shown that in case of a boost-invariant
production process, causality requirements lead to the existence of causally
disconnected spatial production regions. Globally conserved quantum numbers
(charge, strangeness, baryon number) must therefore be conserved within these
regions, which are smaller than the effective overall global volume of a
grand canonical description. This provides a theoretical basis for the 
smaller strangeness conservation volumes just mentioned. In high energy
heavy ion collisions, the superposition of several such volumes from
individual nucleon-nucleon collisions in the same rapidity region is expected
to remove such constraints.

\medskip

\begin{figure}[h]
\centerline{\psfig{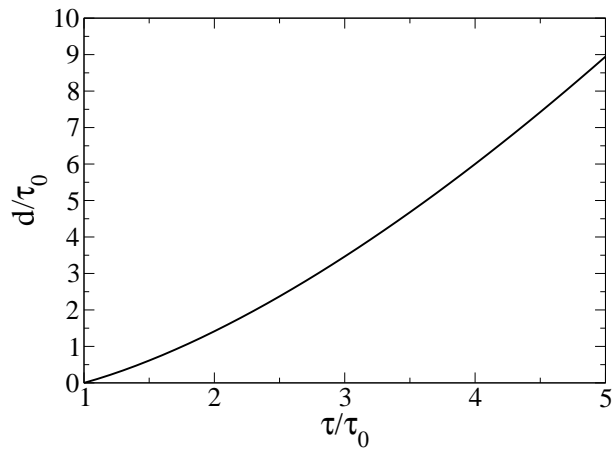}}
\caption{The correlation scale $d$ as function of hadronization time
$\tau_h/\tau_0$}
\label{d-tau}
\end{figure}

In the present paper, we want to quantify these considerations somewhat more,
and also show that at extreme collision energies, even proton-proton
collisions are expected to lead to full chemical equilibrium for strangeness.
The crucial quantity is the size $d$ of the causally connected interaction 
region at the time of hadronization. It was found to be \cite{CS-c}
\be
{d \over \tau_0} =  \sqrt{{\tau_h \over \tau_0}}\left({\tau_h \over \tau_0} 
- 1 \right),
\label{1} 
\ee
where $\tau_0$ and $\tau_h$ denote the equilibration time 
(quark-gluon plasma formation time) and the hadronization time (color 
confinement time), respectively. The resulting variation of the 
correlation scale $d$ is shown in Fig.\ \ref{d-tau}; with the canonical
choice $\tau_0 \sim 1$ fm, that gives the scale in $fm$.

\begin{figure}[h]
\centerline{\psfig{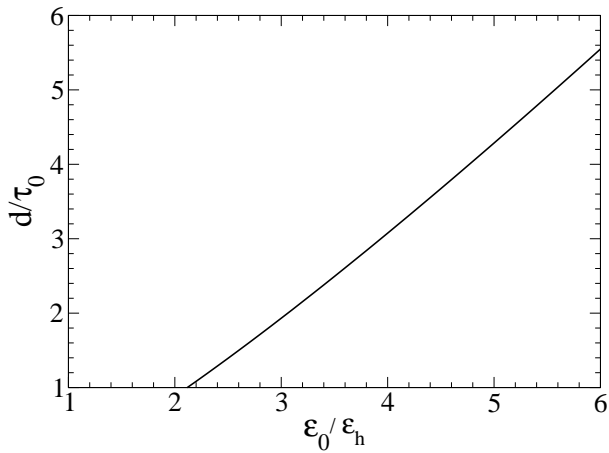}}
\caption{The correlation scale $d$ as function of initial energy density
$\e_0$}
\label{d-epsilon}
\end{figure}

For one-dimensional isentropic expansion,
corresponding to boost-invariant production, the times are related to
the corresponding entropy densities $s$,
\be
s_0 \tau_0 = s_h \tau_h.
\label{2}
\ee
For ideal gas behavior, entropy and energy density are related by
$$
s \sim \e^{3/4,}
\label{3}
$$
so that we then have
$$
{\tau_h \over \tau_0} = \left({\e_0 \over \e_h}\right)^{3/4},
\label{4}
$$
where $\e_0$ is the initial energy density of the collision, and $\e_h$ is
the energy density at the hadronization transition, for which lattice 
calculations give $\e_h \simeq 0.4$ GeV/fm$^3$ \cite{lat-edens,karsch}. 
Combining
this with eq.\ \ref{1}, we obtain in Fig.\ \ref{d-epsilon} the variation
of the scale parameter with the initial energy density $\e_0$.

\vskip0.7cm

\begin{figure}[h]
\centerline{\psfig{file=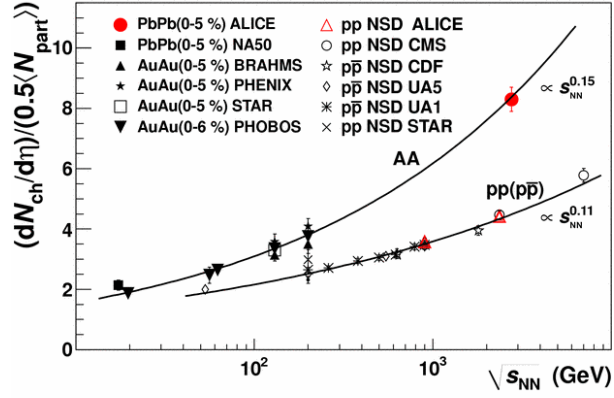,width=8cm}}
\caption{The charged hadron multiplicity per participant in $AA$
and in $pp$ collisions as function of the collision energy 
$\sqrt s$ \cite{multi-data}.}
\label{ppAAmult}
\end{figure}

For the assumed boost-invariant production, the initial energy density 
$\e_0^{AA}$
for $AA$ collisions is given in terms of the resulting hadron production 
through the well-known relation \cite{Bj-edens}
\be
\e_0^{AA} = {m_T \over \tau_0 \pi R_A^2} \left({dN_{AA} \over dy}\right)_0
=   {m_T A^{1/3} \over \tau_0 \pi R_0^2} \left({dN_{AA} \over dy}\right)_0.
\ee
Here $m_T \simeq 0.5$ GeV is the average transverse momentum per hadron,
$R_A \simeq A^{1/3} R_0$  the average nuclear radius, with $R_0=1.25$ fm,
and $(dN_{AA}/dy)_0$ the average hadron multiplicity at central rapidity. We
parametrize the latter quantity as $(dN_{AA}/dy)_0 = A (dN_A / dy)_0$, with
$(dN_A / dy)_0$ denoting one-half of the hadron multiplicity per participant 
in the $AA$ collision. This multiplicity is shown in Fig.\ \ref{ppAAmult}, 
where it is compared to the corresponding quantity in proton-proton 
collisions \cite{multi-data}.
In that case, we have
\be
\e_0^{pp} = {m_T \over \tau_0 \pi R_p^2} \left({dN_{pp} \over dy}\right)_0,
\ee
where $R_p \simeq 0.8$ fm denotes the proton radius. 

\medskip

\medskip

\begin{figure}[h]
\centerline{\psfig{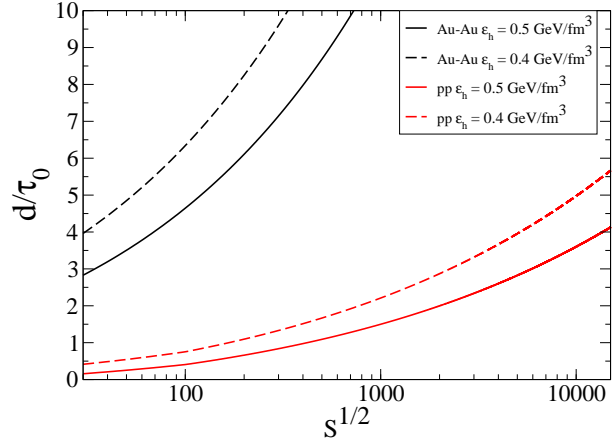}}
\caption{The correlation scale $d$ for $pp$ and $AA$ collisions, for different
input values of $\e_h$}
\label{d-ppAA}
\end{figure}

Using the results of Fig.\ \ref{ppAAmult}, we can now compare the correlation
scale $d(\sqrt s)$ in $AA$ and $pp$ collisions. The result is shown in Fig.\
\ref{d-ppAA}. 

\vskip0.5cm

It is evident that the higher energy density in $AA$ collisions
leads at a given $\sqrt s$ to much larger correlation volumes than that found
for $pp$ interactions. We can use these results to address two issues:
\begin{itemize}
\item{At what incident energy has the correlation scale reached a 
value in $AA$ collisions, for which we should expect grand canonical 
considerations to be valid? An experimental indication is provided by 
the disappearence of strangeness suppression in such collisions. The issue
has also been addressed in theoretical studies \cite{redlich2}.}
\item{At what incident energy does the scale parameter reach a comparable 
value in $pp$ interactions? This would indicate when we should expect 
strangeness suppression to vanish also in such elementary collisions.}
\end{itemize}

\begin{figure}[h]
\hskip4cm{\psfig{file=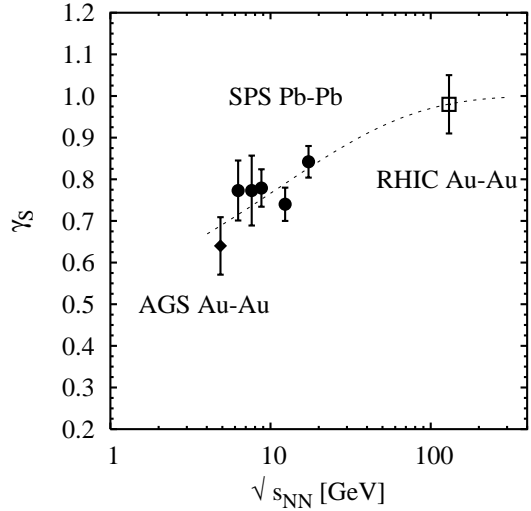,width=10cm}}
\caption{$\gamma_s$ for heavy ion collisions at different energies \cite{mbg}}
\label{gammas}
\end{figure}

To address the first issue, we recall that a parametric way to take 
strangeness suppression into account in the statistical hadronization model
is the introduction of a suppression
factor $\gamma_s$, multiplying by $\gamma_s^n$ the Boltzmann factor of each 
species containing $n$ quarks \cite{LRT}. 
Fits of low energy nuclear collision
data lead to $\gamma_s \simeq 0.6$, increasing with collision energy.
In Fig.\ \ref{gammas}, the behavior of $\gamma_s$ vs.\ $\sqrt{s}$ for 
nucleus-nucleus collisions \cite{mbg} shows that for $\sqrt{s} \simeq 30-40$ 
Gev the strangeness suppression essentially disappears. 

\begin{figure}[h]
\centerline{\psfig{file=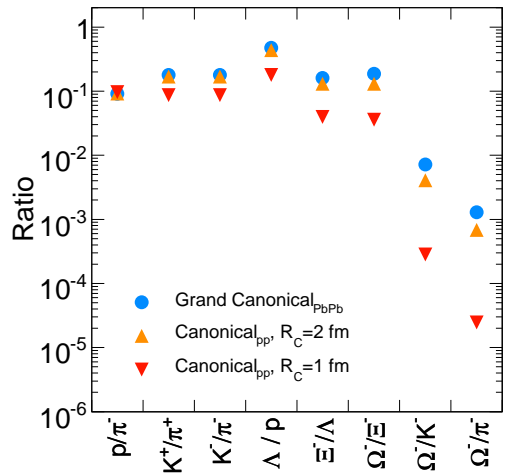,width=7cm}~~~}
\caption {Comparison of canonical and grand-canonical results for different 
size of the strangeness correlation volume \cite{redlich2}}
\label{can-grandcan}
\end{figure}

The mentioned alternative of local strangeness conservation \cite{Red1}
is based on exact conservation (``canonical'' formulation)
combined with a smaller volume of correlation radius $R_c$. It is within
this volume that strangeness must be conserved, and with increasing $R_c$,
one evidently recovers the grand canonical form. As shown in
Fig.\ \ref{can-grandcan} \cite{redlich2}, this occurs for a 
strangeness correlation radius of about $2$ fm; note that $d \simeq
2 R_c$.
 

\begin{figure}[h]
\centerline{\psfig{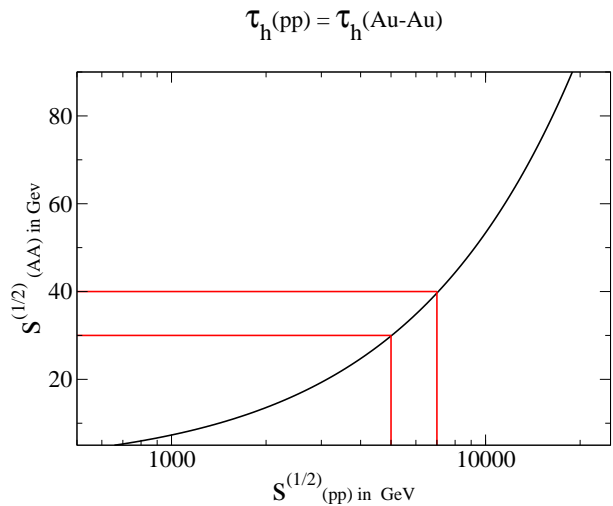}}
\caption{The collision energy values for which the initial energy density
$\e_0$ in $pp$ is equal to that in $AA$}
\label{equaledens}
\end{figure}

For the second issue, we have to determine at which collision energy the
energy density in $pp$ interactions reaches the value at which in AA
collisions strange\-ness suppression vanishes. In Fig.\ \ref{equaledens}
we show the collision energy values in $AA$ and in $pp$ collisions, for
which the energy densities of the two interactions are equal. We note
that to obtain the values at which in $AA$ collisions strangeness suppression
vanishes, $pp$ interaction require collision energies of some 5 - 7 TeV.
It should be emphasized here, however, that a somewhat larger correlation 
volume than that used here remains definitely possible, and that would shift
the necessary $pp$ energy to higher values. The values we have obtained 
just reach the highest presently obtainable LHC energies, and there are 
already first indications showing a considerable increase of strangeness 
production in $pp$ collisions at 900 GeV and at 7 TeV 
\cite{conf-data1,conf-data2}.

\medskip

It would obviously be of great interest to see if high energy $pp$ results
approach the $AA$ results also for the other two indicators mentioned,
quarkonium suppression and jet quenching.

\end{document}